\input epsf.tex
\documentstyle[12pt]{article}
\def\bea{\begin{eqnarray}}
\def\eea{\end{eqnarray}}

\def\beq{\begin{equation}}
\def\eeq{\end{equation}}
\def\ba{\beq\new\begin{array}{c}}
\def\ea{\end{array}\eeq}

\begin{document}
%\begin{center}{\bf Junctions and the Fate of Branes in External
%Fields}
%\end{center}

\setcounter{footnote}0

\setcounter{footnote}0

\begin{flushright}
ITEP/TH-15/99\\
TPI-MINN-99/21 \\
hep-th/9904041
\end{flushright}
\vspace{0.5cm}

\begin{center}
%\hfill ITEP/TH-18/98\\

%\vspace{1.3in}
{\LARGE\bf Junctions and the Fate of Branes in
External Fields}
\date{today}

\bigskip {{\Large A.Gorsky \footnote
{permanent address ; ITEP, Moscow, 117259, B.Cheryomushkinskaya 25}}
\\
Institute for Theoretical Physics, University of Minnesota}

\bigskip{{\Large K.Selivanov } \\
ITEP, Moscow, 117259, B.Cheryomushkinskaya 25}
%$^{\dag}$}
%\\
\end{center}
\bigskip

\begin{abstract}
We discuss the processes of  brane bubble nucleation induced by the
external branes. The quasiclassical solution for the nucleation by
the single external brane has been found in the case when the brane
junctions
are possible. Exponential factor in the production rate
has been calculated. The process induced by
the fundamental or D string in the background of two D3 branes
is analyzed  and
its interpretation from the D3 worldvolume theory viewpoint is
described.
\end{abstract}

\section{Introduction}

During the recent years branes were recognized
as an important ingredient
in description of  the nonperturbative
sector of string and field theories.
Essentially speaking the branes are
multi-dimensional objects having tension
and bearing some quantum numbers
which can be considered as  the
higher-dimensional generalizations of the
electric and magnetic charges (see for instance \cite{polchinski}
as  a review).
In general  the branes are very
natural objects to study, since they are natural generalizations of
the
familiar pointlike particles. Including them into the string
theory considerations has enriched the string
physics as much as Maxwell
theory physics is enriched by adding charges and monopoles.

In the context of string theory the branes are mainly considered as
background for string dynamics.
Some dynamical questions of brane
physics  have also been studied   in the
classical approximation.  Consideration
of the Schwinger-type processes in
external fields involves the simplest
quasiclassical behavior and the paper
is devoted to their analysis.

The spontaneous Schwinger-type processes have previously been
considered in \cite{Teitelboim},
\cite{Porrati}, \cite{dggh}, \cite{emparan},
\cite{MS}{%%Maldacena-Strominger}. Some related classical
solutions to the equations of motion of Born-Infeld (BI)
brane action which support the string like configuration, representing
electrically or magnetically charged
pointlike object on the brane worldvolume,
have been discussed in \cite{cm}{%%Callan-Maldacena},
\cite{gibons}, \cite{Savvidy}.
Recently similar solutions involving string junctions were
found in \cite{ga}.

The basic example of the  brane tunneling phenomena
was analyzed in \cite{Teitelboim} where the brane
creation in the external RR field has been described.
The RR fields  couple to the higher dimensional
generalization of electric charges and like the point particle
Lagrangian includes a term involving the potential one form
integrated over the
worldline of the particle, the $p$-brane action includes the RR
(or NS) $p+1$-form integrated over $p+1$-dimensional worldvolume of
the
brane; a coefficient in front of this term is the higher dimensional
generalization of the electric charge.

We remind that the Schwinger's pair production in the
external electric field is intuitively described as virtual
charge anticharge pair being accelerated by the
external field in the classically
forbidden region until their energy reaches their mass threshold.
The acceleration in the classically
forbidden region is neatly formulated in
terms of imaginary time trajectories, that is, in terms of
trajectories
in Euclidean space which solve the Euler-Lagrange equations obtained
from
the original  Euler-Lagrange equations by the analytical continuation
of time. In the case of uniform external electric field the
corresponding
trajectories are circles of a fixed radius. Analogously, brane
production
in the external
RR fields  is described in terms of the brane world surfaces
which minimize (extremize) the Euclidean  brane action.
In the case of
uniform RR field the corresponding surfaces
describing $p$-brane production
are $p+1$-dimensional spheres (bubbles) of a
given radius. The action
computed on such spheres defines the exponential
factor in the production rate.
The process described can be characterized as a
spontaneous brane production
in the external RR field.

Apart from the spontaneous brane production one can consider {\it
induced}
brane production, that is production of branes in the presence of
some original brane configuration, which is our
main concern here. Since in the
nonperturbative string theory  branes can couple to branes
the external branes can couple to the bubbles and hence can deform
them
changing the brane production rate. The
physics arising is a higher dimensional
generalization of the one discussed in \cite{SV}{%%Selivanov-Voloshin}
(see also consideration
of the case of nonzero temperature \cite{temp}
or density \cite{gk}), where it was shown that in the
particle induced  vacuum decay the bubble is
deformed in the presence of
external particle(s). The key point concerning the induced decay is
the existence of the localized mode of the particle on the kink or
antikink. Due to zero mode the initial
particle can transfer its energy
into the energy of tunneling. The same phenomenon takes place in the
brane bubble creation thanks to the existence of junction
\cite{junction} which substitutes
the particle-kink vertex in the higher dimensional case.
Indeed the initial
brane looses almost all its initial energy after the junction
point and this energy leads to the increase of the nucleation rate.

Hence the crucial point for our further analysis is the existence
of the
string junctions when three (p,q)  strings join in the vertex which
is subject to the charge conservation and zero total tension condition
\cite{junction}. It was shown \cite {dm,ahashimoto}
that the string junction
keeps 1/4 of the original SUSY and can
therefore  be treated classically
so that the generic string networks can be
developed \cite{sen}. More
recently junctions have been promoted to the
M-theory configurations
\cite{M-theory} where the vertex is resolved
smoothly. Note also that
the analogous webs have been elaborated for the 5-branes within
string or M-theory approach \cite{kol}. In what follows we shall
use junctions to find out the explicit
form of the Euclidean solution
corresponding to the induced tunneling.

When considering strings stretched between external D3 branes
interesting interpretation of the induced tunneling emerges. The key
point is that the end of the fundamental (F) string on the D3 brane
can be considered as the point charge \cite{strom}, the
end of the D string as the monopole \cite{diac} and the
end of (p,q) string as the dyon.
Hence any process involving the strings between D3 branes amounts
to the specific process in the SUSY gauge theory
on the D3 worldvolume \cite{witten}.
For instance the processes with the exchange of the quantum
numbers in the bulk have the
interpretation of the nonperturbatively
mediated phenomena involving the generic dyonic states, such as a
well known process of the scattering
of the monopoles to dyons \cite{ah}.
In what follows we shall use
junction configuration with the background D3 branes
\cite{hashimoto} to describe such processes in the
external field.

It is perhaps important that the induced brane
production can be given
a slightly different interpretation - the one of
quantization of the external
branes in an external RR field. The bubbles then describe
contribute to imaginary
part in self-energy (``self-tension'') of the
external branes. Contribution
of such "brane loops" has been discussed
recently for topological strings
in the M theory framework \cite{vafa}.
Furthermore, in terms of the field theory on the inducing brane,
the  external field is a perturbation
of the theory. Therefore the induced
bubble has something to do with the renormalization group flow.

The rest of the paper is organized as follows.
In section 2 we introduce
some necessary background. In section 3 we
discuss our main example -
the induced brane production with one external brane.
The production rate in the leading exponential approximation
is calculated . Section 4 contains
the consideration of the $(p,q)$
strings between D3 branes in
the external field and the
interpretation of the induced tunneling
in terms of the gauge theory on the D3 brane.
Discussion on the related issues and
open questions can be found in Section 5.

\section{Spontaneous brane production}

Consider a $p$-brane omitting the fermionic degrees
of freedom. Its action, essentially, consists of two pieces
\begin{equation}
\label{action}
S=S_{tension}+S_{charge},
\end{equation}}
where $S_{tension}$ is
\begin{equation}
\label{tension}
S_{tension}=T \int_{V} \sqrt{det({\hat G}_{\mu \nu}+
\cal{F}_{\mu \nu})},
\end{equation}
the integration is over the $p+1$-dimensional world-volume $V$
of the brane,
$G_{\mu \nu}$ is the metric induced on the
worldvolume via its embedding into
the target space and the coefficient $T$ is the tension of the brane.
Additional contribution  $\cal{F}_{\mu \nu}=F_{\mu \nu}-B_{\mu \nu}$
is due the U(1) YM field and the pullback of the
background NS two-form B field .
The target space is taken to be the flat
Euclidean one  in the this section,
its metric being
\begin{equation}
\label{metric}
ds^{2}= \Sigma_{i}(dx^{i})^{2}.
\end{equation}
The second term $S_{charge}$ is different
for the fundamental and D branes.
For fundamental brane it has a simple form
\begin{equation}
\label{charge}
S_{charge}=Q \int_{V} {\hat C},
\end{equation}
where ${\hat C}$ is   NS $p+1$ form potential
integrated over the worldvolume of the brane.
For D brane the additional CS term comes
from the RR external fields and
has the structure

\beq
S_{charge}=\int {\exp {\cal{F}}} \wedge \tilde{C},
\eeq
where $\tilde{C}$ collects all relevant RR forms.
In the most interesting
case of D string it includes the excited  two-form and axion fields

\beq
S_{D-str}=\int \tilde{C_2}+\tilde{C_0}\wedge \cal{F}.
\eeq

In external fields the branes are produced in the form of bubbles.
Hence if the brane worldvolume forms a closed manifold (the bubble),
the CS term can
be  rewritten as
\begin{equation}
\label{volume}
S_{charge}= Q \int_{U} {\hat H},
\end{equation}
where $H=dC$ and $U$ is the manifold with the boundary $V$.
In the simplifying assumption that $H$ is a uniform $D=p+2$-form
in the flat Euclidean space of dimension $D$
\footnote{The assumption about
dimension is not restrictive because
the $D$-dimensional Euclidean space can be a subspace of the
target space supporting the flux
of the field $H$; it is energetically
profitable that the bubbles are produced in this subspace and in the
quasiclassical consideration the
rest of the target space is irrelevant.}
the last term reads
\begin{equation}
\label{volume}
S_{charge}={\pm} Q{\Phi} \int_{U} \sqrt{det({\hat G}_{\mu \nu})},
\end{equation}
where ${\Phi}$ is a flux density of the field $H$. The
two possible signs
in Eq.(\ref{volume}) refer to possible orientations of $V$.

Thus, in view of Eqs.(\ref{tension},\ref{volume}),
the brane bubble action
Eq.(\ref{action}) becomes a sum of
surface and volume terms (the latter having
negative coefficient for the relevant bubbles)
\begin{equation}
\label{actionprim}
S=T \int_{V} \sqrt{det({\hat G}_{\mu \nu})}-
 Q{\Phi} \int_{U} \sqrt{det({\hat G}_{\mu \nu})},
\end{equation}
where $Q{\Phi}$ is positive for the relevant bubbles.
Hence there is a competition
of the two terms: the surface term
suppresses small bubbles while the
volume term blows up sufficiently large bubbles.
The one which extremize
the action Eq.(\ref{action}) is the
critical bubble. In the present case
the critical bubble is a $p+1$-dimensional sphere of radius
\begin{equation}
\label{radius}
R=\frac{(p+1)T}{Q{\Phi}}.
\end{equation}

The critical bubble can be obtained in many ways.
We shall now briefly
describe two of them because we believe
each of them is instructive.
Probably the simplest one
is to make the spherically symmetric  anzatz and
to extremize the action
Eq.(\ref{actionprim}) with respect to the radius.
This gives an algebraic
equation giving the critical radius $R$ from
Eq.(\ref{radius}).
The extremal value of the action
Eq.(\ref{actionprim})
\begin{equation}
\label{exponential}
S_{c}=\frac{1}{p+1} {\Omega}_{p+1}TR_{c}^{p+1}
\end{equation}
defines the exponential of the production rate
\begin{equation}
{\Gamma} {\sim} e^{-S_{c}}
\end{equation}
where  ${\Omega}_{p}=\frac{2{\pi}^{\frac{p+1}{2}}}{{\Gamma}(\frac{p+1}{2})}$ is
volume of unit $p$-sphere.
This type of argument was given in \cite{Langer}.

Note that under variation of the form of the bubble
the surface term in the
action Eq.(\ref{actionprim}) gives the so-called ``Laplace pressure''
(or, in other words, trace of the external curvature)
multiplied by tension $T$, while the volume term
gives $-Q{\Phi}$, so in this case one gets the equation
\begin{equation}
\label{pressure}
T(\Sigma_{i} \frac{1}{R_{i}})=Q{\Phi},
\end{equation}
where  $\frac{1}{R_{i}}, i=1, \ldots ,p+1$
are principal curvatures of
the  world-volume of the brane. In the case of spherical symmetry
all $R_{i}$'s are equal and one comes back to Eq.(\ref{radius}).
From this picture it is immediately seen that
for $p=0$ brane the bubble
in any case can only be glued from
arcs of the circle of radius $R$
Eq.(\ref{radius}) \cite{SV}.

The other way, followed in \cite{VKO}, consists
in choosing among the
Euclidean
coordinates a ``time'' coordinate and
assuming the spherical symmetry only
for the remaining ones. Then Eq.(\ref{actionprim}) reduces to
the effective action
\begin{equation}
\label{effective}
S_{eff}=T{\Omega}_{p} {\int}dt {\rho}^{p} \sqrt {1+{\dot {\rho}}^{2}}-
\frac{Q{\Phi}}{p+1}{\Omega}_{p} {\int}dt {\rho}^{p+1}
\end{equation}
where  ${\rho}$ is
the radial coordinate in the spherical coordinate system:
\begin{equation}
\label{spherical}
ds^{2}={\Sigma}_{i}(dx^{i})^{2}=dt^{2}+dr^{2}+r^{2}(d{\Omega}_{p})^{2}
\end{equation}
and $(d{\Omega}_{p})^{2}$ is the metric on the unit $p$ sphere.
Extremization of $S_{eff}$ from Eq.(\ref{effective}) gives an
equations
for ${\rho}(t)$ which defines the critical bubble. One can
use time translation
symmetry of the effective Lagrangian
Eq.(\ref{effective}) to reduce the order
of the equation for ${\rho}(t)$ from second one to the first one.
Namely, the first integral reads
\begin{equation}
\label{energy}
{\cal E}=-\frac{R{\rho}^{p}}{\sqrt{1+{\dot {\rho}}^{2}}}+{\rho}^{p+1},
\end{equation}
where $R$ is as in  Eq.(\ref{radius}).
Since the tunneling proceeds at
$E=0$  Eq.(\ref{energy}) becomes an equation of the sphere of
the radius $R$. The section $t=0$ of the critical bubble is the
configuration which is born in Minkowski space and then  blown
up  by the external field.

\section{One brane induced brane production}

Essentially, the picture of the induced
brane production in the external
field looks as follows. There is an infinite
brane which do not interact
with the external field so asymptotically
its worldvolume is flat.
If the worldvolume of this brane
(we shall call it external) can glue to
(or, end on)
worldvolume of branes which interact
with the external field then the
bubble made of the worldvolume(s) of the
new brane(s) can arise somewhere
in the middle of the world volume of the
external brane. The brane action
computed on such  configuration of branes
(minus action computed on the
configuration which consists of the external brane alone)
defines exponential factor in the brane production rate.
A configuration which emerges (and then blows up)
in Minkowski space consists
of a spacelike slice
of the external brane worldvolume with
the equatorial slice of the bubble
somewhere in the middle of it.

In string theory, the brane worldvolume can
glue to worldvolume of a
brane of the same dimension or to a worldvolume
of a brane of higher
dimension.
In the former case, gluing is possible for the
$(P,Q)$-branes, in the latter case gluing
is possible when the
higher dimensional brane bears an excitation
of internal degrees of freedom,
basically, an excitation of the gauge field living on the brane .
In the present paper we
concentrate on the first case.

In IIB string there are $(P,Q)$ strings (1-branes) and
$(P,Q)$ 5-branes \footnote{Actually there are also $(P,Q)$ 7 branes
but we shall not discuss them later.}. The $(P,Q)$-branes are
branes bearing two types of charges  -
RR and NS ones , correspondingly
there are two types of CS terms in $(P,Q)$-brane action.
As to the tension of the $(P,Q)$-brane, it depends on the charges
in the following way:
\begin{equation}
\label{tensionpq}
T_{P,Q}=T|P+Q{\tau}|,
\end{equation}
where the complex parameter ${\tau}$ encodes axion and dilaton v.e.v.
The week string coupling limit
corresponds to ${\tau} \rightarrow i\infty$
case.
Note that in what follows we shall assume that the RR axion
field is not excited.

From the analysis of  the string junctions
and  (P,Q) webs \cite{sen, kol},
it is known that (P,Q) branes worldvolumes can glue under two
conditions:
1)conservation of the charges;
2)mechanical equilibrium of tension forces.
Thus, e.g., (1,0)-brane couples to (1,1) and (0,-1) branes, and
at ${\tau}=i$ the angle between (1,0)
and (1,1) ones  will be equal to
$3{\pi}/4$,
and the angle between the (1,0) and  (0,-1)
branes  will be equal to ${\pi}/2$.

Obviously, the conditions of gluing  are not locally  affected by the
external field.
What is affected by the external filed
is the shape of the worldvolume
(as seen from Eq.(\ref{pressure})). However, for any  brane of a
given configuration of charges one
can choose such a linear combination of
RR and NS external fields that the brane does not feel it in the
leading approximation. Below we
assume that the external field is of NS type, while the external brane
is a D-brane, that is with charges of $(L,0)$ type, and hence it does
not
interact with the external field.  $(L,0)$-brane can couple to
$(P,Q)$ and $(L-P,-Q)$ branes and
worldvolumes of the latter are curved
by  the external NS field. Locally
they are glued according to the same
condition 1) and 2) above, that is
the angle ${\alpha}_{(L,0)(P,Q)}$
between $(L,0)$ and $(P,Q)$ is defined according to
\begin{equation}
\label{alpha}
cos {\alpha}_{(L,0)(P,Q)}= -\frac{P+Q Re{\tau}}{|P+Q{\tau}|}
\end{equation}
and analogously  the angle  ${\alpha}_{(L,0)(R-P,-Q)}$ is given by
\begin{equation}
\label{alpha1}
cos {\alpha}_{(L,0)(R-P,-Q)}= -\frac{L-P-Q Re{\tau}}{|P-L+Q{\tau}|}.
\end{equation}
Note that there is an essential difference here from the (P,Q)-webs
case - the internal branes are not flat, their worldvolumes form  sort
of ``caps'' (see Fig.1) which are glued to the
external brane worldvolume at the
angles defined by Eqs.(\ref{alpha}),(\ref{alpha1}), with  one of the
caps
is glued at the
angle defined by Eq.(\ref{alpha}) from above and  the other one -
at the angle defined by Eq.(\ref{alpha1}) from below.

\begin{figure}[t]
\hspace*{5cm}
\epsfxsize=7cm
\epsfbox{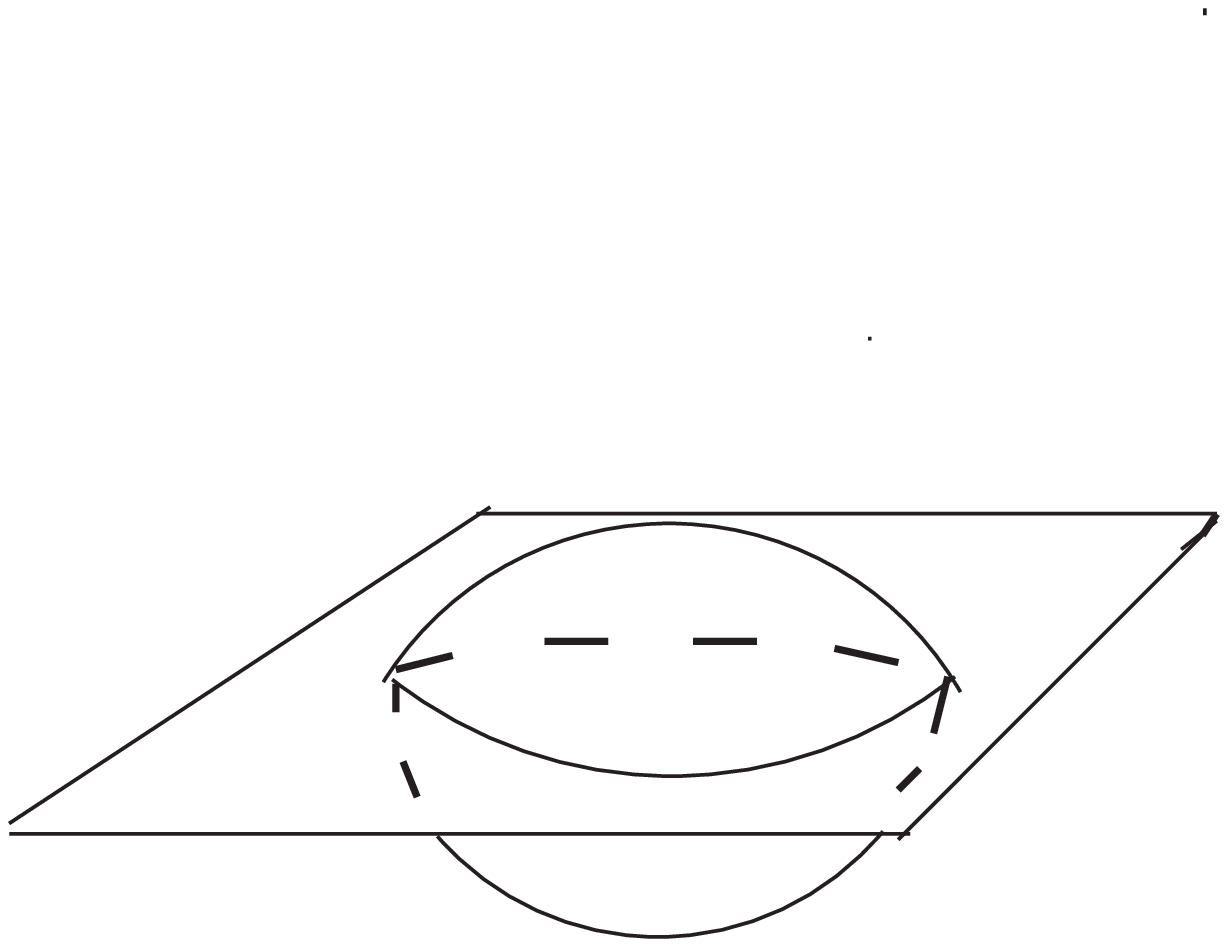}
\caption[x]{Stringy induced Euclidean solution.}
\label{IA}
\end{figure}

Let us turn to the discussion of the gauge field contribution.
Thanks to a point junction Gauss law
on the worldsheet D string theory
has to be modified
since the F string inserts the point source.
Due to the discontinuity of
the electric field $\Delta E=g$,where g is the IIB string coupling,
The $A_0$ component has to be piecewise
linear and according to BPS condition
has to be
correlated with one of the scalars representing
the transverse fluctuation of the string \cite{dm,ahashimoto}.
This condition is fulfilled
at the junction and is equivalent to the total zero tension condition.

In the case considered instead of the
point like junction point we have
junction circle around the bubble. The Gauss law now reads  as
\beq
divE=g(\delta (x - x_0(t))- \delta (x + x_0(t))
\eeq
where $\rho$ denotes the radius of the circle in the (x,t) plane
(the plane of the external  string),
$x_{0}^2+t^2=\rho^2$. Solution to the Gauss law constraint
provides the discontinuity of the
electric field. Once again BPS condition
amounts to the total zero tension along
the circle which is consistent
with the equations of motion to the gauge field and scalar. Let us
emphasize that we use the BI action
to describe the D string. Introducing
the canonical momentum for the gauge field $ \Pi=\frac{\delta L_{BI}}
{\delta E}$ and performing the
Legendre transform to the Hamiltonian
description one immediately recognizes
the action of the string with
tension $T_{P,Q}$ since the canonical momentum
along the "caps" is constant due to the
equations of motion.

To find the shape of the caps it is most
convenient to turn the second of the
two possibilities described at the
end of section 2. A coordinate $z$,
orthogonal to the
external brane worldvolume (see footnote in section 2),
will be considered as ``time'' in the effective problem.
\footnote{Perhaps, it is worth to
stress that this is unphysical time,
which is
convenient to describe the shape of
the caps, because the caps are obviously
spherically symmetric in the rest of
coordinates. The physical time - the
one in which the brane production should be interpreted - is, in
fact,  one of the coordinates along the external brane.}
With this choice, the effective action reads
\begin{equation}
\label{action1}
S=S_{external}+S_{(P,Q)}+S_{(L-P,-Q)},
\end{equation}
where $S_{external}$ includes only tension
term for the external brane and
$S_{(P,Q)}$ and $S_{(L-P,-Q)}$ include both the tension and CS terms
(cf. Eq.(\ref{effective}):
\begin{equation}
\label{action1}
S_{(P,Q)}=T_{(P,Q)}{\Omega}_{p} {\int}_{0}^{t_{+}}dt {\rho}^{p}
\sqrt {1+{\dot {\rho}}^{2}} -
\frac{Q{\Phi}}{p+1} {\Omega}_{p}  {\int}_{0}^{t_{+}}dt {\rho}^{p+1}
\end{equation}
and
\begin{equation}
\label{action2}
S_{(L-P,-Q)}=T_{(L-P,-Q)}{\Omega}_{p} {\int}_{t_{-}}^{0}dt {\rho}^{p}
\sqrt {1+{\dot {\rho}}^{2}} -
\frac{Q{\Phi}}{p+1} {\Omega}_{p}  {\int}_{t_{-}}^{0}dt {\rho}^{p+1},
\end{equation}
where tensions $T_{(P,Q)}$ and $T_{(L-P,-Q)}$ are defined in
Eq.(\ref{tensionpq}), and $t_{+}, {\rho}=0$ ( $t_{-}, {\rho}=0$) is a
top
(bottom) point of the upper (lower) cap.
Notice that the volume terms in Eq.(\ref{action1}) and
Eq.(\ref{action2})
have the same  coefficient which is traced
back to the charge conservation.

Instead of Eq.(\ref{energy}) of  section 2 we now have two
first integrals - the one for the
upper cap and the other - for the lower cap.
Moreover it is clear that the first integral
constant ($E$) vanishes, so the caps are segments  of spheres.
The radii of the spheres are those
of the spheres in the spontaneous brane
production:
\begin{equation}
\label{radius1}
R_{(P,Q)}=\frac{(p+1)T|P+Q{\tau}|}{Q{\Phi}}
\end{equation}
for the $(P,Q)$-cap and
\begin{equation}
\label{radius2}
R_{(L-P,-Q)}=\frac{(p+1)T|P-L+Q{\tau}|}{Q{\Phi}}
\end{equation}
for the $(L-P,-Q)$ cap. Given the radii Eqs.(\ref{radius1}),
(\ref{radius2}),
the segments are uniquely defined by the angles
Eqs.(\ref{alpha}),(\ref{alpha1}).
To verify that the caps fit into a bubble one
has to check that the $p$-spheres on the
boundaries of the segments have the
same radius which is   indeed the case.
To find the exponential of the brane production rate one should take
the value of the effective action Eq.(\ref{action1})
on the configuration
described and subtract from it action
computed on the configuration which
consists of the external brane alone.

The contribution from the external
string in Eq.(\ref{action1})is, of course,
infinite. However, the decay rate is defined by the difference
between the critical value of the action from  Eq.(\ref{action1})
and the value of the action when only
flat external brane (without any bubble)
is present. This difference is finite and defines the following rate
of
the induced brane production
\begin{eqnarray}
\label{rate}
{\Gamma}=exp \{- \frac{1}{(p+1)(p+2)}{\Omega}_{p+1}
 \frac{ {\left ((p+1)|P+Q{\tau}|T \right )}^{p+2}}{(Q{\Phi})^{p+1}}
\frac{1}{\pi} arcsin \left | \frac{QIm{\tau}}{P+Q{\tau}}\right|
\nonumber\\
- \frac{1}{(p+1)(p+2)}{\Omega}_{p+1}
 \frac{ {\left ((p+1)|P-L+Q{\tau}|T \right )}^{p+2}}{(Q{\Phi})^{p+1}}
\frac{1}{\pi} arcsin \left | \frac{QIm{\tau}}{P-L+Q{\tau}}\right|
\nonumber\\
+ \frac{1}{(p+1)(p+2)}
{\Omega}_{p}L
\frac{ {\left ((p+1)T \right )}^{p+2}}{({\Phi})^{p+1}}
|Im{\tau}|^{p+1}\}
\end{eqnarray}
${\Gamma}$ can equally be considered as
an imaginary part of the $(L,0)$
tension in the external NS field.

The effective contribution from the external string
comes from the difference between the action of the external
brane without and with the bubble
solution and in the string case
reads
\begin{equation}
S_{ext}=-T_{1,0}\pi \rho^2,
\end{equation}
where $\rho$ is the intersection radius.
Let us emphasize that the
contribution $S_{ext}$ can be treated
differently as the contribution
due to the Wilson loop along the matching
circle. This interpretation
is in perfect accord  with the Gauss law
if we attribute the Wilson loop
to the effective "charges". Note once
again that the effects due
to the classical gauge fields on the caps are taken into account
by tensions of the branes.

It is in order here to compare the rates of the spontaneous and
induced brane nucleation. Of course, this comparison cannot be taken
literally since the final states are different. However it is
instructive to identify the enhancement factors for the induced processes
in different regimes.

Let us recall the situation for the false vacuum decay induced by
a particle of mass $m$ in (1+1) dimensions. The relevant parameter
is the ratio ${\epsilon}=m/2{\mu}$ where $\mu$ is the kink mass
(the tension of the bubble). At small $m$ the external particle doesn't
disturb classical bubble solution significantly. Therefore the
enhancement factor is just exponential of the particle action
inside the bubble, $exp^{2mR_{0}}$, where $R_{0}={\mu}/{\epsilon}$
is the radius of the unperturbed bubble. Opposite limit
corresponds the so-called ``sphaleron'' region
$\epsilon \sim 1$ where the exponential factor disappears since the
initial particle can decay into the kink-antikink pair without
external field. Note however that in the sphaleron region large quantum
corrections make inapplicable the saddle point approximation.

Now turn to the brane induced processes. The relevant parameter
analogous to the parameter $\epsilon$ above is  the parameter $\tau$
defining the ratio of tensions of $D$ and $NS$ branes. The region of the
slightly disturbed bubble corresponds to the limit
$Im{\tau} \rightarrow \infty$ and the enhancement factor in this limit
is again given by exponential of the external brane action
inside the bubble,
\begin{equation}
\label{rate1}
{\Gamma}_{enh}=exp\{\frac{1}{(p+1)}
{\Omega}_{p}L
\frac{ {\left ((p+1)T \right )}^{p+2}}{({\Phi})^{p+1}}
|Im{\tau}|^{p+1}\}
\end{equation}

The ``sphaleron'' region is where the initial brane can pass into
a pair of branes without external field. The corresponding
``line of marginal stability''( where the tension
of the initial brane equals sum of the tensions of pair
of the produced branes)
is reached at $Im \tau =0$. Discussion of  quantum
corrections to the  ``line of marginal stability'' is beyond
the scope of the paper.
Eq.(\ref{rate}) predicts that there is no suppression of the induced
brane production at this line.
However, in analogy with the particle induced vacuum decay,
Eq.(\ref{rate}) cannot be trusted in the sphaleron region because
of strong string corrections.

Let us now comment on the choice of the initial D-brane state. The
subtlety concerns the possibility to have
the initial brane at rest. If
we consider the initial D-brane one can wonder about the dependence
of the general BI action on the NS B field
and therefore the corresponding
interaction of D brane. To handle this issue for a string
we can suppose that gauge fields are not
excited in the initial state
and there is no axion $\tilde{C_0}$ field at
all. However in the generic
case the noncommutative geometry induced by NS
B field is involved
hence the noncommutative BI  Lagrangian actually governs the dynamics.

It is natural to discuss the more general
case when two external branes
are involved in the induced process.
Let us  consider  two $(1,0)$ branes as external
branes (see fig.2) inducing the tunneling.
In this case
the bubble describing branes production
consists of two caps, one  above
and one  below , and
a barrel in the middle. One of the caps
has charges $(1,1)$, the other one -
$(1,-1)$, and the barrel - $(0,1)$.
The external branes (``legs''),
the caps, and the
barrels are glued at the angles defined by Eq.(\ref{alpha}).
%for example, for the angles indicated in fig.2 we have
%\begin{eqnarray}
%\label{beta}
%cos{\alpha}=-\frac{1+{\tau}}{|1+{\tau}|}\nonumber\\
%cos{\beta}=\frac{\tau}{\tau|}
%\end{eqnarray}
Notice, that
in vicinity of the bubble the external
branes are not parallel and not flat.
The effective action consists of various pieces - those for the  legs,
for the caps and for the barrel. One can see that the caps are again
segments of spheres, while for
the barrel the first integral - analogue
of Eq.(\ref{energy}) is not zero. Its external
curvature has two different
eigenvalues. What concerns to the legs, there is no ``volume term''
in the effective action for them which allows
them to spread to infinity,
and the first integral is again nonzero which makes them curve.
There is no principal difficulties in
describing the bubble in this case,
but the formulae are not as transparent
as in the one brane induced
case, and we will not bother the reader by them.

Now, after describing all these induced bubbles,
we would like to come back
to the spontaneous brane creation. It is now
clear, that in addition to the
basic case of round bubble made out of one
brane, there are other bubbles
of branes spontaneously arising
in the external field. For example,
one can have a bubble glued from two curved branes bearing
$(P.Q)$ and $(-P-L,-Q)$ charges
(``caps'') and one flat brane
in the middle with $(L,0)$ charge .
The production rate of such bubbles is described by Eq.(\ref{rate})
with the change of $L$ to $-L$.

\begin{figure}[t]
\hspace*{5cm}
\epsfxsize=7cm
\epsfbox{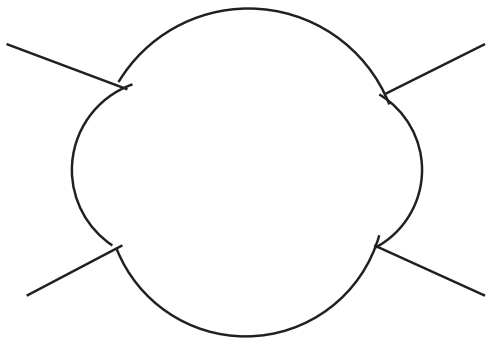}
\caption[x]{Time slice of the Euclidean solution with two external
branes .}
\label{IA}
\end{figure}

\section{Nucleation of brane bubbles in D3 background}

So far we considered the one-brane
induced process in the flat space.
It is interesting to consider the string induced process in the
background
of two D3 branes far apart from each other.
Namely consider the initial F or D string stretched
between two D3 branes along some direction.
In this case the ends of the
string on D3 branes represent monopoles \cite{diac}
(D string) or charges \cite{strom}
(F string) in the 4d SYM theory
on the D3 brane worldvolume.
If distance between D3 branes  corresponding to the vacuum
expectation
value of the scalar in 4d theory is large
we can ignore the metric
deformation due to D3 branes and consider the
perturbative domain in 4d theory. The configuration
under consideration is presented
at Fig.3.

In the external B field the process looks as follows.
The initial D string
creates a bubble in the Euclidean
space which then evolves in Minkowski
space. Since the simplest decay mode
for D string is to F string and
(1,-1) string, from the point of
view of the 4d observer on D3 brane
the process looks like the
exponentially suppressed decay of
monopole to charge and dyon
in the external field during the finite time
determined by the bubble
action.

The time-evolution for the process looks as follows.
The critical (turning point) configuration is the one
which arises in the Minkowski region. Then the external field
accelerates the (1,1) and (-1,0) strings and at some time
size of the bubble becomes comparable with the distance
between D3 branes. At this moment in terms of the $SU(2)$
theory on D3 branes (1,1) dyon and (-1,0) are created.

Note that we can believe such consideration
if the vacuum expectation
value of the scalar is much larger then
the critical radius, otherwise
the $AdS_{5}\times S^5$ structure of the near horizon
D3 brane metric

\beq
ds^2=H^{-\frac{1}{2}}dx_{||}^2 + H^{\frac{1}{2}}
(dr^2+ r^2d\Omega_{5}^2)
\eeq
\beq
H=1+ \frac{4\pi g\alpha^{'2}}{r^4},
\eeq
where $X_{||}$ are four coordinates
along the D3 worldvolume and
$d\Omega_{5}^2$ is the five-sphere metric
has to be taken into account.

\begin{figure}[t]
\hspace*{5cm}
\epsfxsize=7cm
\epsfbox{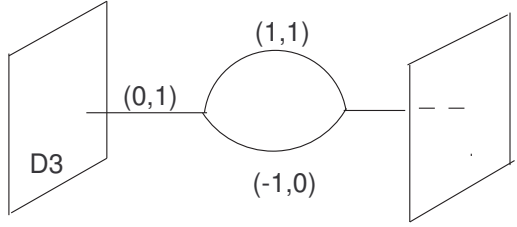}
\caption[x]{Time slice of the Euclidean solution
for the induced process in D3 background}
\label{IA}
\end{figure}

Let us note that the process looks  somewhat unexpected from the
point of view of 4d action on D3 brane where  two-form NS field
enter through the BI term while the RR one through CS one.
Since the BI action supports the F and D string excitations
\cite{cm,gibons} we
can claim that these excitations are
unstable in the RR or NS two form
background respectively.
Having this
in mind we can interpret the process from the D3 brane observer
viewpoint as the
"instanton" like monopole decay in the NS B field (for D string)
or charge decay in the RR B field (for F string).
Both processes are exponentially
suppressed due to the dyon production
rate by the Euclidean tunneling
exponent.

The decay rate can be presented in the form
\beq
\Gamma =f_{mink}\exp(-S_{Eucl}),
\eeq
where $S_{Eucl}$ has been found in the previous section while the
$f_{Mink}$ stands for the Minkowski part
of evolution till the moment when
dyonic strings touch D3 branes.
Minkowski contribution to the action is purely imaginary
\beq
S_{Mink}=i\int p_{Mink}(r)dr,
\eeq
where $p_{Mink}$ is the canonical momentum conjugate to the
the radial variable. The subtle point is the integration region.
In principle there are two different situations; in the first case
the effective potential for the radius has no any additional
extremum in the D3 brane background and the contribution of
the Minkowski part of the evolution doesn't play the important role.
However one can't exclude the possibility
that there is such extremum
and the Minkowski evolution allows the
finite motion. In this case
the amplitude develops the poles related to the energy levels
for the radial variable in the
Minkowski region. The possibility
of such scenario requires the
careful analysis of all gravitational
effects and deserves further investigation.

Let us turn to the case of two D strings. Without external field
we have two monopoles with the nontrivial
moduli space. The brane configuration
fits perfectly Nahm description of the moduli space \cite{diac}.
Nahm equations
themselves
\begin{equation}
D_sT_i=-\epsilon_{ijk}[T_j,T_k]
\end{equation}
describe the condition of the BPS
invariance of the configuration and
s is identified with the direction along the strings.
Matrixes $T_i$ correspond to the positions of D string ends on the
D3 branes.
Solution to the Nahm equations with
the proper boundary conditions amounts to the moduli space of the
two-monopole configuration and provides the hyperkahler metric.

External field yields the deformation of the picture. Consider the
SU(2) theory on the worldvolume of D strings. The
external field enters the Lagrangian
through the BI action and hence deforms
Nahm equations.
The new B field dependent metric can
in principle  be found in a standard way
using the construction of the spectral
curve from the solution of the
Nahm equation or using the hyperkahler structure \cite{ah}.

From the generic point of view the
state of two monopoles can be represented
by the point in the moduli space while their slow relative motion
is governed by the geodesic motion in the hyperkahler metric.
In the external B field the unstable
submanifold in the moduli space
is  developed. Indeed assuming that the bubble size is smaller
than the distance between D3
branes we have to conclude that any
point at this submanifold is unstable due to the decay
of two static monopoles to the
dyons due to the bubble solution.
The instability reflects the
presence of the negative mode which
results in the exponentially suppressed process.

One more interesting
comment is in order here. The Nahm equations are the monopole
generalization of the ADHM construction
for instantons. It is also known
that the monopole can be thought of as the  chain of
instantons. In the brane picture
above this can be recognized viewing
the D1 string as the bound state of the infinite number of D(-1)
branes representing instantons. Since the distance from D3 brane
corresponds to the instanton size \cite{bg}
we can conjecture that the bubble
creation can be considered as the nontrivial deformation of the
instantons of the sizes  related with the radius of the
bubble .

The case of the nontrivial momenta of monopoles
looks more complicated. Even in the absence of the external
fields there are some nonperturbative
phenomena in this case. We can
mention the scattering at the right
angle in the forward collision or the
transition to the dyons at the
generic kinematics \cite{ah}. In the external
field the picture is even more rich. The  inspection
of the matching conditions for
the bubble solution with the forward collision
of two D string in the B field shows that there is
a possibility to have different kinematics
at the final state therefore besides
the scattering at the right angle
there is the exponentially suppressed processes
with generic angle kinematics.
Moreover there is a plenty of
possibilities in the generic
kinematics for the nonperturbative
phenomena. For instance there
is the process of transition of
the pair of initial monopoles or
charges into the generic state of
several dyons through the more complicated "bubble" type solution.

\section{Discussion}

In this paper we developed
quasiclassical approach to a higher dimensional generalization
of the induced tunneling processes.
Utilizing the existence of the brane
junction configuration the explicit
quasiclassical tunneling exponent has
been calculated. Our main example
involves the stringy induced amplitudes
but the 5 brane case is treated along the
same routing. The solutions
found above can be combined with the
string-like solutions to the BI equations
of motion to get the generic webs in the
external fields. Similar
arguments can be applied to  more
generic processes analogous to
the charge and monopole decays mentioned
above. Let us also remark that
in the usual field theory framework
there are slightly different processes
similar to the induced false vacuum decay via bubble creation,
for instance the processes
yielding the baryon number non-conservation.
The corresponding
nonperturative solution involves
instanton-antiinstanton pair and the collective
coordinate relevant for tunneling is
the Chern number. It seems that
the results of the paper might be
useful  for the quantitative description of
their higher dimensional analogies.

Not much can be said about the quantum corrections to the induced
processes. It is not clear  if the quantum correction can be
reduced to the geometric characteristics of the solution. The only
remark available concerns the effect of the quantum fluctuations
of the gauge fields
to the stringy induced process since the relevant 2d YM action is
almost topological and depends only on the area of the
manifold which the YM theory is
defined on. In our case the relevant manifold
has the topology of the sphere therefore the expectation value of
the Wilson loop in 2d YM theory is of interest. The Wilson loop
expectation can be formulated in term of group characters \cite{gt}
and even calculated at the large N limit \cite{dk}. It appears
that the dependence on the area  exhibits some phase
transition behavior but for large
areas it manifests the standard area law and therefore actually
yields  the tension renormalization.

Let us mention that the proper playground for the theory with
the external NS B field is the
noncommutative geometry. Having in mind that
the rate of the process of the creation of branes depends
nonanalitically on the B field
one can expect that the partition function of the theory
can manifest some singular structure similar to the singularities
at the complex coupling plane in the usual field theory due to
instantons. In particular, since the ratio of the process can be
considered as imaginary (nonanalitical in $B$ !)
part of the brane tension, one can expect a change of the notion
of BPS states in the noncommutative case, as compared to \cite{SK}.
We shall be back to this subject elsewhere \cite{GS}.

Let us turn now to the gauge theories
on the worldvolume on the emerging branes. The situation is
most transparent in the case of 5 brane webs. The particular
configuration of 5 branes with some "external 5 brane legs"
amounts to SU(N) 5d gauge theory on the worldvolume
of N 5 branes  stretched between the
external ones \cite{kol}. In the external
field some brane production is possible which means the process of
nonperturbative change of the rank of the gauge group whose rate
depends nonanalitically on the external field. Moreover the
brane diagram itself can be treated as a kind of Euclidean solution
corresponding to the nonperturbative phenomena from the point of view
of the gauge theories on the {\it external } 5 branes. Finally let
us mention that the Minkowski evolution of the 5 brane bubbles
acquires the meaning of some RG flow in the theory on their
worldvolumes.

It would be interesting to discuss the case
when temperature or density
play the role of the inducing factor. In the temperature case it is
necessary to find the Euclidean solution periodic in the Euclidean
time. The temperature fixes the size of the matching p-sphere and
therefore gives rise to the action on the solution. The formulae
in this case are similar with the substitution of the temperature
instead of the tension of the external brane. It is not clear how
the density case has to be treated since the chemical potential
for the state with branes has to be
defined properly. Let us also note
that the induced process can be considered in the AdS spaces
generalizing the consideration in \cite{MS}.

We are thankful to A.Mikhailov, M.Shifman and M.Voloshin for the
useful
discussions. A.G. thanks to TPI at University of Minnesota were the
paper
was completed for the hospitality. The work was supported in part by
grants
INTAS-96-0482   and RFBR 97-02-16131
(A.G) and INTAS-97-0103 (K.S).

\end{document}